# Getting Genetic Ancestry Right for Science and Society

*There is a scientific and ethical imperative to embrace a multidimensional, continuous view of ancestry and move away from continental ancestry categories*

Anna C. F. Lewis[1], Santiago J. Molina[2], Paul S Appelbaum[3,4], Bege Dauda[5], Anna Di Rienzo[6], Agustin Fuentes[7], Stephanie M. Fullerton[8], Nanibaa' A. Garrison[9,10,11], Nayanika Ghosh[12], Evelynn M. Hammonds[12,13], David S. Jones[12,14], Eimear E. Kenny[15,16], Peter Kraft[17], Sandra S.-J. Lee[18], Madelyn Mauro[1], John Novembre[6], Aaron Panofsky[9,19,20], Mashaal Sohail[21], Benjamin M. Neale[22,23,24*], Danielle S. Allen[1*]

[1] Edmond J Safra Center for Ethics, Harvard University, Cambridge, MA, USA
[2] Department of Sociology, Northwestern University, Evanston, IL, USA
[3] Columbia University Vagelos College of Physicians and Surgeons, New York, NY, USA
[4] New York State Psychiatric Institute, New York, NY, USA
[5] Center for Global Genomics and Health Equity, University of Pennsylvania, Philadelphia, PA, USA
[6] Department of Human Genetics, University of Chicago, Chicago, IL, USA
[7] Department of Anthropology, Princeton University, Princeton, NJ, USA
[8] Department of Bioethics & Humanities, University of Washington School of Medicine, Seattle, WA, USA
[9] Institute for Society & Genetics, University of California, Los Angeles, CA, USA
[10] Institute for Precision Health, David Geffen School of Medicine, University of California, Los Angeles, CA, USA
[11] Division of General Internal Medicine & Health Services Research, David Geffen School of Medicine, University of California, Los Angeles, CA, USA
[12] Department of the History of Science, Harvard University, Cambridge, MA, USA
[13] Hutchins Center for African and African American Research, Harvard University, Cambridge, MA, USA
[14] Department of Global Health and Social Medicine, Harvard Medical School, Boston, MA, USA
[15] Institute for Genomic Health, Icahn School of Medicine at Mount Sinai, New York, NY, USA
[16] Department of Medicine and Genetics and Genomic Sciences, Icahn School of Medicine at Mount Sinai, New York, NY, USA
[17] Program in Genetic Epidemiology and Statistical Genetics, Harvard T.H. Chan School of Public Health, Boston, MA, USA
[18] Division of Ethics, Columbia University, New York, NY, USA
[19] Department of Public Policy, University of California, Los Angeles, CA, USA
[20] Department of Sociology, University of California, Los Angeles, CA, USA
[21] Centro de Ciencias Genómicas (CCG), Universidad Nacional Autónoma de México (UNAM), Cuernavaca, Morelos, México
[22] Broad Institute of Harvard and MIT, Cambridge, MA, USA
[23] Analytic and Translational Genetics Unit, Massachusetts General Hospital, Boston, MA, USA
[24] Center for Human Genetic Research, Massachusetts General Hospital, Boston, MA, USA
* These authors contributed equally to this work

Glaring health disparities have reinvigorated debate about the relevance of race to health. This includes how race should and should not be used as a variable in research and biomedicine (*1*). Following a long history of race being treated as a biological variable, there is now broad agreement that it is instead a socio-political construct, meaning that how we classify people by race is a product of historically contingent social, economic, and political processes. For example, in the US immigrants from southern and eastern Europe were only put into the category "white" on the census during the twentieth century (*2*); American Indian and Alaska Native peoples are defined sociopolitically (*3*). As such, social scientists and others have argued that the strongest case for using race is best limited to tracking the impact of racism on health outcomes, rather than as a proxy for anything biological (*4*).

Currently, many academic and healthcare institutions have been re-examining their use of race and its links to racism, and stating intentions about how race should be used in the future. One common proposal is to use genetic concepts — in particular genetic ancestry and population categories — as a replacement for race (*5*). However, this proposal risks retaining one of the most problematic aspects of race—an essentialist link to biology—by allowing genetic ancestry categories to stand in its place.

Genetic ancestry is certainly relevant in many different disciplines. Beyond statistical and population geneticists, it is of relevance to epidemiologists, public health practitioners, physicians and patients. In particular, it has renewed relevance for the clinical application of genetic technology because the accuracy of genetic risk scores is influenced by ancestry (*6*). Genetic ancestry and population categories are also relevant to the general public, as demonstrated by the tens of millions of individuals who have paid for ancestry reports from consumer companies such as 23andMe.

Across these different domains, a dominant understanding of genetic ancestry is based on continent of origin. Within genetics research, continental ancestry categories have become the most common type of group label (*7*). Similarly, consumer genetics reports give customers a report with data based on a percentage of these continental groups from which an individual can trace their "ancestry."

Racial classifications have often taken continents as boundaries between human groups; thus it is not surprising that racial categories and continental ancestry categories are often confounded. Whenever continental ancestry categories are used, the risk is high that a misconception of race as biological will re-enter through the backdoor (*8*). It becomes straightforward to connect race-based differences in health status to genetic ancestry through the use of continental ancestry categories, for example, associating black-white racial disparities in the impact of COVID-19 in the US with allele frequency differences between those of European and African ancestry (*9*). This makes it easy to use genetic explanations to explain race-based differences in health status, even in the presence of much more plausible socio-environmental explanations. Attempts to isolate possible biological contributions to race-based health disparities using the percentage of a particular continental ancestry category risk circularity. This is because if the continental ancestry categories correlate with race (which many maintain, and use this to justify this use of race in medicine (*10*)), they also correlate with racism and other non-biological factors.

In addition to the conflation with racial categories, there are at least three additional interconnected problems with the use of continental ancestry categories. Collectively, these illustrate that current practices around ancestry estimation and reporting limit the accuracy and reliability of claims being made by researchers about human genetic difference.

First, there are deep-seated ambiguities about what genetic ancestry actually is. There is no widely agreed upon definition. There are many statistical methodologies across sub-fields of genetics and genomics whose outputs are framed as "genetic ancestry" (*11*), see Figure 1. Some of these attempt to approximate the subset of paths through a family tree via which DNA has been inherited. But many do not even attempt to do this, and are better thought of as estimates of *genetic similarity* between individuals in a dataset (*11*), rather than as genetic ancestry. This applies to the commonly used technique of assessing relative position in principal component space, i.e. a space that captures the major axes of variation in a dataset. Framing this instead as genetic similarity helps bring to the fore the question "similar compared to whom?", and hence the crucial role of which individuals are selected in the analysis.

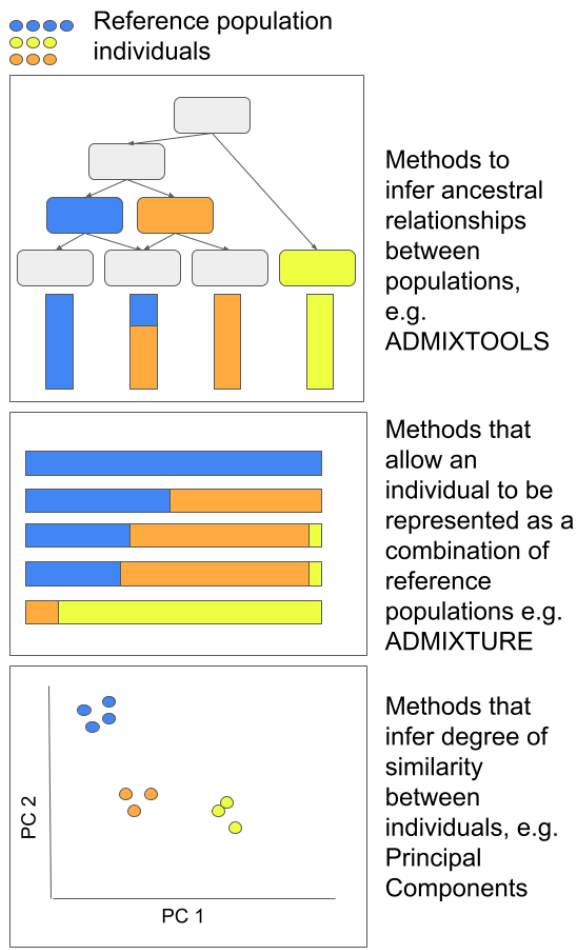

*Figure 1. What is meant by genetic ancestry? What is meant by "genetic ancestry" is ambiguous. The outputs of several statistical methodologies are framed as revealing genetic ancestry; many of these are better thought of as capturing genetic similarity. All measures depend sensitively on input data, and the contextualization of all results depends on how the input data is labeled.*

Second, imposing any categories on genetic ancestry fails to adequately capture human genetic diversity and what we know of human demographic history. A standard way to visualize genetic similarity is via a principal component plot. If individuals from the most commonly used reference populations are graphed in this manner, distinct clusters roughly representing continental categories are visible. These groups represent an emergent space in terms of genetic ancestral diversity. Indeed, a prominent early result was that genetic ancestry was remarkably concordant with self-identified continental ancestry (*12*). But if people are sampled differently, such as individuals living in New York City, it becomes clear how impoverished this view of a structure of distinct clusters is (see Figure 2). More specifically, the clearly separated clusters of reference population individuals, corresponding to different continental groups, merge into a background of continuous genetic variation. This is consistent with what we know of human demographic history, in which mass migration and constant mixing across groups have been the norm throughout our species' past. It is worth emphasizing that the impact of these histories leads to different structures of genetic variation in different parts of the world.

The continuous, i.e. category free, nature of genetic variation has been appreciated since at least the introduction of the intra-group cline (a gradation in a given characteristic of a species over geographical space) in the 1930s by evolutionary biologist and eugenicist Julian Huxley and, in the context of humans, since biological anthropologist Frank B. Livingstone's work in the 1960s. But newly assembled datasets, such as that visualized in Figure 2, illustrate just how inappropriate use of discrete continental categories can be, particularly when information framed as genetic ancestry can potentially influence medical care.

Recognizing the growing existence of "admixed individuals" — typically defined as those who have recent ancestry from more than one population — does not escape the notion of continental ancestry categories but rather compounds it, because an individual is considered a mixture of these broad continental groups. In addition, we have learnt from population history that continental groups are extraordinarily diverse and that humans have been constantly mixing within and across continents. We need to be able to describe every human; the only way to do this is to adopt a fully continuous view of ancestry.

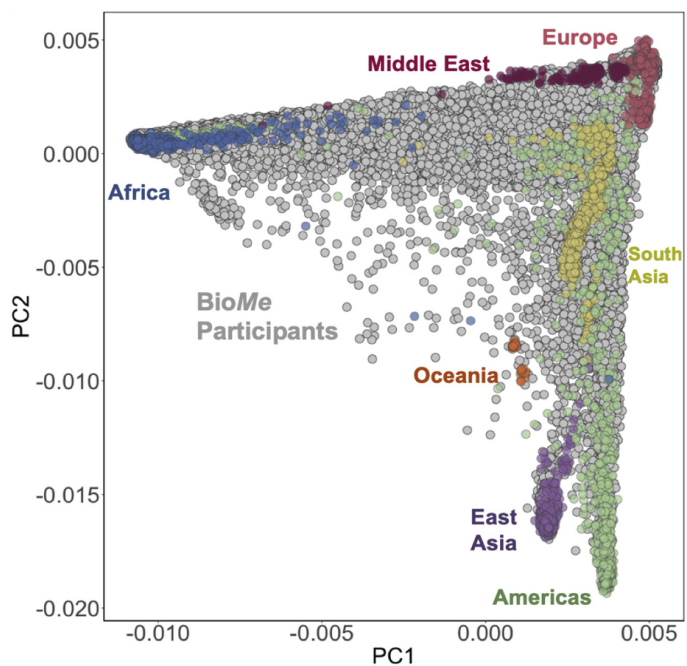

*Figure 2. The continuous, category free, nature of genetic variation. Reproduced from Belbin et al, [Towards a fine-scale population health monitoring system.](#) This image shows individuals projected onto the first two principal components of genetic similarity. Colored dots are N=4149 reference panel individuals from 87 populations representing ancestry from 7 continental or subcontinental regions. Gray dots are N=31705 participants from BioMe, a diverse biobank based in New York City. Clearly delineated continental ancestry categories, the islands of color, are shown to be a by-product of sampling strategy. They are not reflective of the diversity in this real-world dataset, made evident by the continuous sea of gray.*

Third, the use of continental ancestry categories oversimplifies complex human history into a snapshot. There is no one answer to "what is my ancestry?" because the answer depends on the time horizon. We each have ancestors from every generation in our species' past. With advances in ancient DNA and in population genetics, a contemporary human genome can increasingly give us visibility into the chronologically layered ancestral record for that person. Yet this historical notion of genetic ancestry is flattened when just one set of categories is used. In the case of continental ancestry categories, their use reflects the assumption that at some specific point in time humans were mostly divided into homogeneous groups by the natural geographical barriers between continents. This is a gross oversimplification of human history. It also obscures other time slices where different categories would be relevant. For example ~50,000 years ago Homo Sapiens and Neanderthal categories, or ~5,000 years ago "Steppe-related", "European" hunter-gatherer and "Near Eastern" farmer categories in Europe (*13*), or ~500 years ago when waves of migration and the slave trade were forging new patterns of human genetic diversity in the Americas. Ancestry categories from other time points may be of medical relevance today (*14*).

Given these issues with the use of continental ancestry categories, and the particular danger of their conflation with racial groups, why are they so often used? In order to produce the largest sample sizes possible and hence be powered to detect smaller effect sizes, current methods of data aggregation often rely on label harmonization approaches, whereby finer-grained categories are replaced by continental categories. Some statistical geneticists find that using these continental groupings represents a way to control for false positive genetic associations while maintaining statistical power to detect the associations in the first place. Science is reductive, and a model that uses these simple categories has been useful in the process of understanding human genetic diversity. But all models have their legitimate domains of application and limits, and a much more complex set of models should be the norm across a wide variety of use cases. This is particularly important because while human genetics falls under the biological sciences, it is in fact a science at the intersection of several disciplines, including history, demography, anthropology and sociology. Even if the limitations of models used are well understood by statistical and population geneticists, others may take the models to be descriptive of realities rather than recognizing that they merely formalize approximations and estimates, using reductive categories to do so. Hence one of the risks of using these categories is that others may interpret them as true natural kinds, which is inaccurate. Instead, they are heuristics permitting the answering of very narrow sorts of questions.

While there may be particular situations where approximating the underlying distribution with a categorical variable is unavoidable, there are other situations where this approximation may not be necessary at all (*15*). In the cases where genetic ancestry categories can be avoided, they should be avoided. Because of the association of continental ancestry categories with racial groupings, this is particularly important for continental categories.

An individual researcher's use of continental ancestry categories is not in and of itself racist, but the cumulative impact of this practice has led to and sustains racism. Typological thinking about human difference has had damaging social consequences. Continued reliance on continental ancestry categories contributes to failures of inference, miscommunication between fields, and reported findings that are rooted in reductive and limited ways of understanding human difference. These are likely to exacerbate medical stereotypes about individuals and groups, contribute to health disparities rather than addressing them, and reify biological (mis)understandings of race as biological. The solution will require addressing the issues with how ancestry is conceptualized and used across the entire biomedical research ecosystem, not just within each subfield of biological science. This will involve the development and operationalization, and hence widespread use, of a more complex notion of ancestry — one which disambiguates what is meant by ancestry, that wherever possible does not treat it as a categorical variable, and treats ancestry as reflecting a historical process.

To arrive at this more complex notion of ancestry a solid empirical understanding of how and why different fields use and operationalize the concept is needed. To ensure this more complex notion of ancestry is then used in practice, systems-level change is needed. New computational tools and data structures will be required. Methods to harmonize data that do not involve substituting finer-grained categories with continental categories will be needed. Educational materials will need to be developed for scientists and physicians. Scientists of all stripes who engage in research that employs biological categories for humans should not work in isolation but as part of interdisciplinary teams, ideally including engagement with impacted communities. In support of these efforts journal editors should set standards, professional societies should publish best practices, and funders should carefully consider which research agendas they will support. It is paramount that as these organizations rightly critique the use of race as a biological variable, that use of continental ancestry categories does not become the new default.

Adoption of a more complex notion of ancestry should in turn continue to inform the research agenda in population and statistical genetics and in ancient DNA research. It is in these fields, the home turf of the concept of genetic ancestry, that change in practice may have the largest overall impact. These changes are a prerequisite to any research that looks for connections between genetics and health disparities. More generally, with a more complex notion of ancestry that reflects continuous variation and historical depth, we can start to pave the way for a science that reflects the complex histories of human groups, including the power dynamics between them.

Sloppily-made connections between race, biology, health and assumptions about the genetic variation between and within continental groupings have done much damage. Many scientists now invoke genetic ancestry as the legitimate biological component of group difference. To make the most of this window of opportunity to move away from race as a biological variable, we would urge the adoption of a multidimensional and continuous conceptualization of ancestry, free wherever possible of population categories, and not relying on continental labels that bear striking resemblance to prior racist groups.

**Acknowledgements**
*Funding:* NIMH administrative supplement 5000747-5500001474 to 3R37MH107649-06S1
*Author Contributions:* ACFL: Writing – original draft; BMN and DA: Supervision; All authors: Conceptualization and Writing – review & editing.
*Competing interests:* ACFL owns stock in Fabric Genomics; EK has received personal fees from Regeneron Pharmaceuticals, 23&Me, and Illumina, and serves on the advisory boards for Encompass Biosciences and Galateo Bio; BMN is a member of the scientific advisory board at Deep Genomics and RBNC Therapeutics, Member of the scientific advisory committee at Milken and a consultant for Camp4 Therapeutics and Merck.